\documentclass[a4paper]{article}

\usepackage[utf8]{inputenc}
\usepackage[english]{babel}
\usepackage{amsmath}
\usepackage{graphicx}
\usepackage{url}
\usepackage{setspace} 
\usepackage[T1]{fontenc}
\usepackage{float}
\usepackage[dvipsnames]{xcolor}
\usepackage[style=numeric, sorting=none]{biblatex}
\usepackage[hidelinks]{hyperref}
\usepackage{csquotes}

\title{
\textbf{Multiclass MRI Brain Tumor Segmentation using 3D Attention-based U-Net}
}

\author{Maryann M. Gitonga \\ \href{mailto:maryann.mwendwa@strathmore.edu}{maryann.mwendwa@strathmore.edu} \\ Strathmore University, Nairobi, Kenya}

\date{\today}

\begin{document}
\maketitle

\begin{abstract}
This paper proposes a 3D attention-based U-Net architecture for multi-region segmentation of brain tumors using a single stacked multi-modal volume created by combining three non-native MRI volumes. The attention mechanism added to the decoder side of the U-Net helps to improve segmentation accuracy by de-emphasizing healthy tissues and accentuating malignant tissues, resulting in better generalization power and reduced computational resources. The method is trained and evaluated on the BraTS 2021 Task 1 dataset, and demonstrates improvement of accuracy over other approaches. My findings suggest that the proposed approach has potential to enhance brain tumor segmentation using multi-modal MRI data, contributing to better understanding and diagnosis of brain diseases. This work highlights the importance of combining multiple imaging modalities and incorporating attention mechanisms for improved accuracy in brain tumor segmentation.


\end{abstract}

\section{Introduction}

Glioma is a common malignant brain tumor that originates from glial cells in the brain and spinal cord. Gliomas are aggressive, and the median survival time for glioma patients is about 12 months \cite{ranjbarzadeh2021brain}. Early detection of these tumors is critical, and MRI is a primary tool used for this purpose. MRI provides high spatial resolution anatomical information and different sequences such as T1-weighted, T2-weighted, T1-weighted contrast-enhanced, and T2 Fluid Attenuated Inversion Recovery, which highlight different tumor characteristics \cite{diaz2021deep}. Accurate annotation and segmentation of tumor borders are essential for tumor diagnosis. However, manual segmentation is costly, time-consuming, and prone to human error, especially in cases where tumors have varying intensities and shapes in different sub-regions \cite{bakas_keyvan}.

\section{Literature Review}
Consequently, deep learning techniques have transformed brain tumor segmentation from feature-driven to data-driven. Two types of deep learning algorithms, Convolutional Neural Network(CNN)-based and Fully Convolutional Network(FCN)-based, are used in brain tumor segmentation. Havaei \textit{et al.} proposed a multi-path CNN network, InputCascadeCNN, which uses variable size convolution kernels to extract context features and has both local and global routes \cite{havaei2017brain}. The model attained a Dice coefficient of 0.81 for the complete segmentation on the BraTS 2013 dataset. Myronenko proposed a 3D CNN-based approach using a shared decoder and a variational auto-encoder branch for regularization. The encoder of the network extracts features of the images and the decoder reconstructs the dense segmentation masks for the scan. The variational auto-encoder branch reconstructs the input image into itself and is used only during training to regularize the shared decoder.  The model attained an average Dice coefficient of 0.82 on the BraTS 2018 dataset \cite{myronenko20193d}. Bukhari \textit{et al.} proposed the 3D U-Net which has a contracting path for capturing context information and an expanding path for ensuring precise location, which greatly improves the performance of medical picture segmentation task. This state-of-the-art model attained a Dice coefficient of approximately 0.92 on the BraTS 2021 dataset\cite{talha2021e1d3}. Lin \textit{et al.} incorporated a feature pyramid module into the U-Net architecture to combine multi-scale semantic and location information. The solution had shortened distance between the output layers and deep features by bottom-up path aggregation to try and reduce the noise in the segmentation. The efficient feature pyramid was used to improve mask prediction using fewer resources to complete the feature pyramid effect. The model attained a Dice coefficient of 0.80 on the BraTS 2017 and BraTS 2018 datasets\cite{lin2021path}. Jun \textit{et al.} introduced a nn Unet architecture which included an encoder and decoder composed of convolutions, normalization, and skip connections, with deep supervision added to all but the two lowest resolutions in the decoder. It attained an average Dice coefficient of 0.90 on the BraTS 2021 dataset \cite{ma2022nnunet}.\newline

\noindent
However, these techniques still take up significant time and resources in training and evaluation to get good results. Additionally, small-scale tumors are difficult to accurately segment due to decreased picture dimension during downsampling. In this paper, I test the 3D attention-based U-Net network \cite{oktay2018attention} with the Dice Coefficient and Tversky Loss Function as metrics, to improve the segmentation accuracy. \cite{singh2022brain} applied the same network but applied Hausdorff Distance and augmented the dataset using the Positive Mining technique. I apply the proposed method to the BraTS (Brain Tumor Segmentation) 2021 Dataset provided by Medical Image Computing and Computer-Assisted Intervention (MICCAI). To provide a richer spatial information in the input as well as enable one-time segmentation, the three modalities out of the four are combined into one. This is because the native modality (T1) highlights the healthy anatomy of the brain and not the tumor regions \cite{md_2023}. This results to a 4D input of dimensions $3M \times L \times W \times S$, where $M$ is the modality, $L$ is the length of the scan, $W$ is the width of the scan and $S$ is the number of slices in each volume. This allows the model to put more focus on the regions of interest (region showing a potential tumor).

\section{Methodology}

\subsection{3D Attention U-Net}

\begin{figure}[H]
\centering
\includegraphics[scale=0.5]{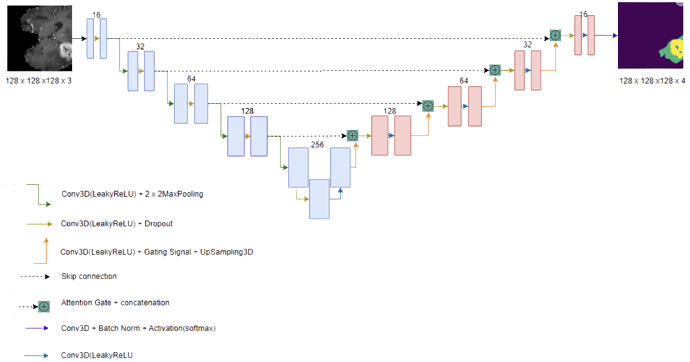}
\caption{3D Attention U-Net Architecture}
\label{fig:unet}
\end{figure}

The proposed architecture (cf. Figure \ref{fig:unet}) uses the U-Net architecture \cite{ronneberger2015u}, which employs a contracting path to down-sample image dimensions and an expanding path to up-sample while retaining spatial information through skip connections. Attention modules are utilized at the skip connections to highlight relevant activations and suppress those in irrelevant regions during training \cite{oktay2018attention}. The soft attention module, which is differentiable and essential for back-propagation, is used, consisting of two sub-modules: the channel attention module and spatial attention module. The former selects important feature maps, while the latter identifies important regions within the feature maps, and both are used to take full advantage of the architecture. 3D attention gates are introduced to generate 3D channel and spatial attention by utilizing 3D inter-channel and inter-spatial feature relationships.

\begin{figure}[H]
\centering
\includegraphics[scale=0.3]{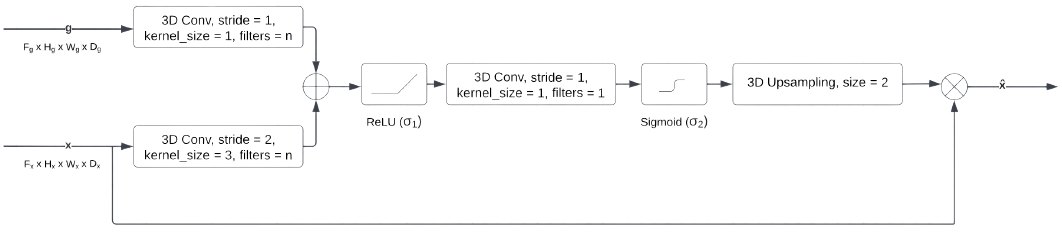}
\caption{Visual representation of 3D Attention mechanism}
\label{fig:attention}
\end{figure}

The attention mechanism (cf. Figure \ref{fig:attention}) consists of an attention gate that takes in two input vectors, x and g. g is acquired from the lower part of the network and represents better features and x comes from early layers and represent better spatial information. To align the dimensions of the two vectors, x undergoes a strided 3D convolution while g undergoes a 3D convolution with number of filters = $F_g$. The two vectors are summed element-wise, with aligned weights becoming larger while unaligned weights become relatively smaller. The resultant vector undergoes a ReLU activation layer and a 1x1x1 convolution that reduces the dimensions to 1xHxWxD. A sigmoid layer is applied to scale the vector between 0 and 1, generating attention coefficients that indicate the relevance of the features. The attention coefficients are upsampled to the original dimensions of vector x and multiplied element-wise with vector x. The resultant scaled vector is passed along in the skip connection. This mechanism helps to increase the sensitivity of the network to small but important details and reduce the impact of noisy or irrelevant features. The overall network is able to learn better and more discriminative features which improves on its accuracy and efficiency.

\section{Experiments}

\subsection{Dataset and Pre-processing}
The BraTS 2021 Dataset, which consists of 1400 cases of multi-parametric MRI (mpMRI) scans with expert neuro-radiologists' ground truth annotations, was used for this project. The dataset provides mpMRI scans in NIfTI format and includes native (T1), post-contrast T1-weighted (T1CE), T2-weighted (T2), and T2 Fluid Attenuated Inversion Recovery (T2-FLAIR) volumes, along with manually annotated GD-enhancing tumor, peritumoral edematous/invaded tissue, necrotic tumor core, and normal tissue. 

\noindent
The scans in each folder were transformed by scaling and translating the features using the MinMax Scaler to shrink the features within the range of 0 to 1 \cite{shaheen2020minmaxscaler}. The combined MRI scan was generated by merging the 3 volumes of each brain scan to form a 4D array of 3(modalities) x length x width x number of slices. This provides richer spatial information for one-time segmentation. The fourth volume (native modality) was left out because the scan highlights the healthy tissues of the brain \cite{md_2023} which does not majorly contribute to the segmentation of the tumor regions. The combined scan and corresponding mask were then cropped to remove useless blank regions, reducing bias and focusing on the important parts of the volume. The pre-processed combined scan and mask were saved as numpy arrays, with the mask features converted to class values (labels) 0, 1, 2, and 3. Masks with a segmented region less than 1\% were excluded to retain only significant feature representation for the segmented regions. The resulting dataset after pre-processing had about 1200 cases of tumor without any additional in-house data. The dataset was divided into three sets: the train, test, and validation dataset in the ratio 6:2:2 respectively.

\subsection{Implementation Details}
During the training of a segmentation model, several hyper-parameters were investigated, including batch size, learning rate, epochs, activation function, dropout rate, metric functions, and loss functions. To prevent overfitting, the dropout approach and batch normalization were utilized for model regularization. The dropout rate was distributed across the layers between the range of 0.1 to 0.3 in both the encoder and decoder modules. To fit the data into memory, a batch size of 2 was utilized, which was the maximum allowable limit based on the GPU specifications obtained for this research. This small batch size also offered a regularizing effect, resulting in lower generalization error \cite{brownlee2019control}. The Adam optimizer with a learning rate of $1.0 \times 10^{-4}$ was utilized for weight updates \cite{shi2022improved}. A pixel-wise softmax activation function was employed in the last layer of the model. The Dice Coefficient (Equation \ref{eq:dice_coef}) was utilized as an evaluation metric for both training and testing phases. It calculates the ratio between the intersection and the union of the segmented and ground truth regions, focusing only on the segmentation classes and not the background class. The pixels are classified as True Positive ($TP$), False Negative ($FN$) and False Positive ($FP$).
\begin{equation}
D = \frac{2TP}{2TP + FN + FP}
\label{eq:dice_coef}
\end{equation}
The Tversky Loss \cite{jadon2020survey} (Equation \ref{eq:tversky_loss}), based on the Tversky Index (Equation \ref{eq:tversky_index}) where $ \alpha = 0.7, \beta = 0.3$ \cite{salehi2017tversky}), was used for both training and testing phases.
\begin{equation}
TI = \frac{TP}{TP + \alpha FN + \beta FP}
\label{eq:tversky_index}
\end{equation}
\begin{equation}
TL= 1 - TI
\label{eq:tversky_loss}
\end{equation}
This loss function is a generalized approach to handle class imbalance issues, resulting in a better balance between precision and recall during model training. ReLU was used in the first trial of training and evaluation and in the second trial, the activation function was changed to LeakyReLU to prevent the dying ReLU problem. \cite{xu2020reluplex}.

\section{Results}

\begin{figure}[H]
\centering
\includegraphics[scale=0.15]{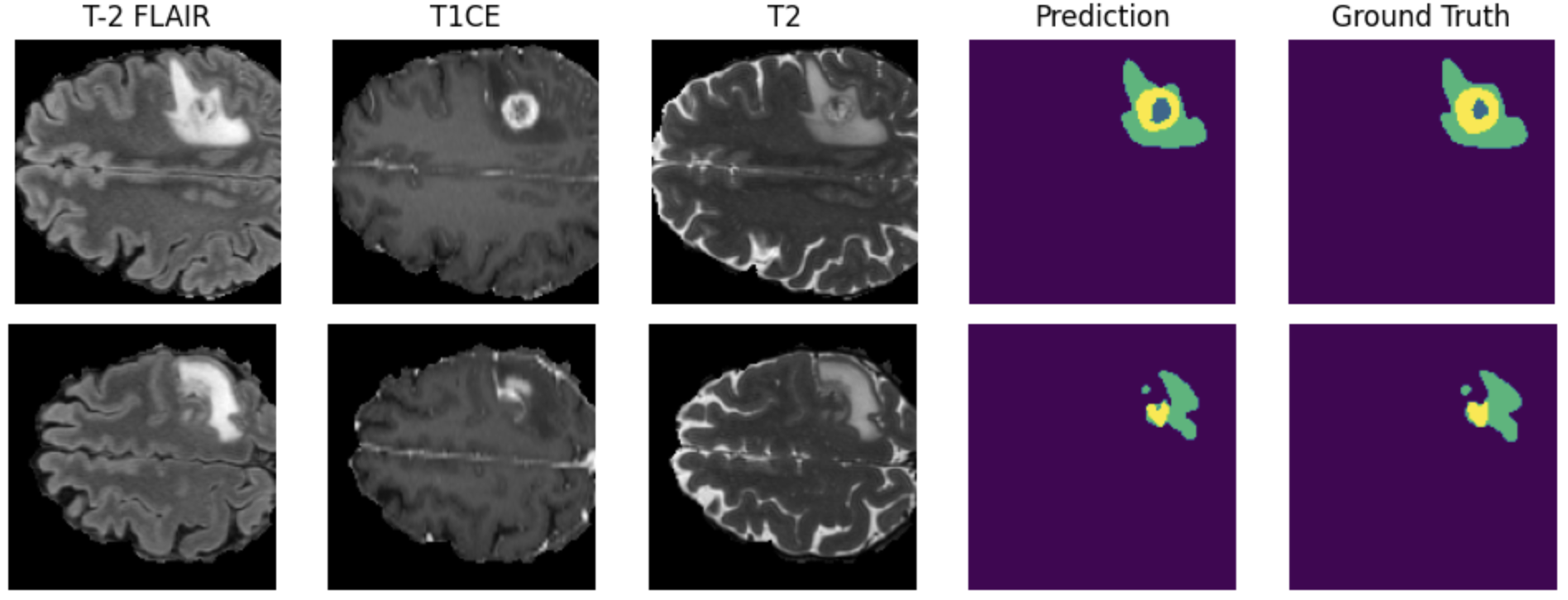}
\caption{ Predictions and Ground truths visualized from a combined scan generated from 3 modalities: T2-FLAIR, T1 and T1CE from the validation set from BraTS 2021. The annotation can be interpreted as \textcolor{Blue}{necrosis}, \textcolor{Goldenrod}{edema (invaded tissue)} and \textcolor{Green}{enhancing tumor}.}
\label{fig:prediction}
\end{figure}

The network was implemented in Tensorflow and trained it on NVIDIA Tesla
V100 32GB GPU. Results for Dice Coefficient and Tversky Loss metrics evaluation on the validation and testing datasets are presented in Table \ref{table:results}. The developed model achieved promising results in brain tumor segmentation, with the best performance attained during the second trial at the 127th epoch. Table \ref{table:comparison} shows the dice coefficients attained by other models from different studies in comparison to the developed model. The use of Dice Coefficient and Tversky Loss metrics evaluation on the validation and testing datasets demonstrated the model's effectiveness. The visualization of the testing dataset is as shown in Figure \ref{fig:prediction}. The model's ability to accurately delineate the tumor and its sub-regions from the input stack of three volumes (T2-FLAIR, T1CE, and T2) helps to create an effective treatment plan based on the nature of the tumor sub-regions observed from the error-proof segmentations obtained.

\begin{table}[H]
\begin{tabular}{p{3.5cm}|p{3.5cm}|p{3.5cm}}
 \hline
 Trial & Dice Coefficient & Tversky Loss\\
 \hline\hline
\end{tabular}
\centering{Validation Dataset (BraTS 2021)} \\
\begin{tabular}{p{3.5cm}|p{3.5cm}|p{3.5cm}} 
 \hline\hline
 Trial 1 (epoch = 75) & 0.9430 & 0.0570 \\ 
 \hline
 Trial 2 (epoch = 127) & \textbf{0.9562} & \textbf{0.0438} \\
 \hline\hline
\end{tabular}
Testing Dataset (BraTS 2021) \\
\begin{tabular}{p{3.5cm}|p{3.5cm}|p{3.5cm}} 
 \hline\hline
 Trial 2 & \textbf{0.9864} & \textbf{0.0136} \\ 
 \hline
\end{tabular}
 \caption{Dice Coefficient and Tversky Loss metrics evaluation on the validation and testing datasets.}
 \label{table:results}
\end{table}

\begin{table}[H]
\begin{tabular}{p{4cm}|p{3.5cm}|p{3cm}}
 \hline
 Model & Dataset & Dice Coefficient\\
\hline\hline
\end{tabular}
\begin{tabular}{p{4cm}|p{3.5cm}|p{3cm}} 
InputCascadeCNN \cite{havaei2017brain} & BraTS 2013 & 0.81 \\ 
 \hline
 Encoder-Decoder with a variational auto-encoder branch \cite{myronenko20193d} & BraTS 2018 & 0.82 \\
 \hline
3D U-Net \cite{talha2021e1d3} & BraTS 2021 & 0.92 \\
 \hline
 Path aggregation U-Net \cite{lin2021path}  & BraTS 2017, BraTS 2018  & 0.80 \\
 \hline
3D Attention U-Net \cite{islam2020brain}  & BraTS 2019  & 0.86 \\
 \hline
 Nn U-Net \cite{ma2022nnunet}  & BraTS 2021  & 0.90 \\
 \hline
  3D Attention U-Net*  & BraTS 2021  & \textbf{0.98} \\
 \hline
\end{tabular}
\caption{Dice Coefficient metric comparison with models from other studies. The developed model is marked with an *.}
\label{table:comparison}
\end{table}

\noindent
This method has significant implications for early brain tumor detection, which is crucial for effective treatment and ultimately saving lives. With tumors being one of the leading causes of mortality worldwide, the model's output is critical in forecasting the tumor's aggressiveness and the patient's survival early enough, allowing for the best chance for successful treatment. The developed solution helps facilitate accurate and effective medical diagnostics by optimizing computational resources consumed on irrelevant areas on the images and facilitating better generalization of the network used for the task.

\section{Conclusion}
In this paper, the model implements a 3D attention mechanism in a 3D U-Net network to improve model sensitivity and accuracy to foreground pixels without requiring significant computation overhead by progressively suppressing features responses in irrelevant background regions. The model performs segmentation on a stack of 3 modalities of the MRI scan, in their original format (NIfTI) to attain richer feature representation during segmentation. This work clearly exhibits the significance of the 3D attention mechanism in multi-class segmentation with limited computational resources. The significance of stacking the modalities into one array is also demonstrated: providing better feature representation in the input and facilitation of one-time segmentation of the multi-modal scans. This solution in its entirety contributes to the development of accurate, extensive delineation tools for brain tumors, allowing the physicians to develop effective treatment plans for the patients based on the nature of the tumor sub-regions observed from error-proof segmentations obtained.

\printbibliography[title=References]
\end{document}